\documentclass[aps,prl,showpacs,twocolumn,superscriptaddress]{revtex4}
\usepackage{amsmath}
\usepackage{amssymb}
\usepackage{epsfig}
\usepackage{graphicx,amsmath}
\begin{document}

\title{Tetragonal tungsten bronze compounds: relaxor vs mixed ferroelectric -
dipole glass behavior}

\author{V.A. Stephanovich}\homepage{http://draco.uni.opole.pl/~stefan/VStephanovichDossier.html} \email{stef@math.uni.opole.pl}
\affiliation{Opole University, Institute of Physics, Opole, 45-052, Poland}
\begin{abstract}
We demonstrate that recent experimental data (E. Castel et al J.Phys. Cond. Mat. {\bf 21} (2009), 452201)
on tungsten bronze compound (TBC) Ba$_2$Pr$_x$Nd$_{1-x}$FeNb$_4$O$_{15}$
can be well explained in our model predicting a crossover from ferroelectric ($x=0$)
to orientational (dipole) glass ($x=1$), rather then relaxor, behavior. We show, that
since a "classical" perovskite relaxor like Pb(Mn$_{1/3}$ Nb$_{2/3}$)O$_3$ is never a ferroelectric, the presence of ferroelectric hysteresis loops in TBC shows that this substance actually transits from ferroelectric to orientational glass phase with $x$ growth. To describe the above crossover theoretically, we use the simple replica-symmetric
solution for disordered Ising model.
\end{abstract}
\pacs{77.80.Bh, 77.84.Lf, 77.84.Dy} \maketitle

Disordered dielectrics and ferroelectrics for a long time have been
attracting much attention of scientists due to additional (governed by disorder)
possibility to control their physical properties which might
be useful for applications. The disordered dielectrics can be in principle
divided on two classes. One of them comprises the compounds like
KTaO$_3$:Li (KTL), Nb (KTN), Na, which undergo a crossover between ferroelectric
and orientational (dipole) glass phases \cite{hoh}, while the other class may be related to so-called relaxor
ferroelectrics. Latter substances, being actually never ferroelectrics \cite{smagr,vax08}, belong
to perovskite family. Their general formulas are Pb(B$_{1/2}$B$'_{1/2}$)O$_3$ (so-called 1:1 family), Pb(B$_{1/3}$B$'_{2/3}$)O$_3$ (so-called 1:2 family) with B-ion being Mn, Zn, Sc, Nb or Ta.  For example, such materials as
Pb(Sc $_{1/2}$ Nb$_{1/2}$ )O$_3$ (PSN), Pb(Sc$_{1/2}$ Ta$_{1/2}$)O$_3$ (PST) (1:1 family) or Pb(Mn$_{1/3}$ Nb$_{2/3}$)O$_3$(PMN), Pb(Zn$_{1/3}$ Nb$_{2/3}$)O$_3$ (PZN) (1:2 family) are considered in the literature to be "classical" relaxors.

The main reason for above relaxors not to become ferroelectrics is the nonstoichometry of the position of the B ion in their perovskite structure. This nonstoichiometry actually "spoils" the phonon spectrum
inherent in perovskite structure, destroying the ferroelectrically important soft
mode (see Ref. \cite{vax09} and references therein). Latter fact manifests itself in many observable quantities of relaxors, to name a few, the smearing of ferroelectric phase transition and the appearance
of additional low-temperature peaks in dielectric spectra obeying Vogel-Fulcher law \cite{vie}.
There are, however, differences between relaxors and disordered dielectrics like KTL.
The major difference is that since above relaxors do not exhibit macroscopic ferroelectricity, they
never have the ferroelectric hysteresis, while the substances like KTL exhibit it in their ferroelectric
and mixed ferroglass phases, see e.g. Ref. \cite{boa}.
As tungsten bronze compound (TBC) Ba$_2$Pr$_x$Nd$_{1-x}$FeNb$_4$O$_{15}$
exhibits ferroelectric hysteresis, which disappears with growth of Pr content $x$, it rather shows the
crossower between ferroelectric and dipole glass behavior similar to that in KTL family.

In this paper, on the base of analysis of the experimental facts about TBC \cite{mag}, we come to conclusion that
this substance demonstrates mixed ferroelectric- orientational glass rather then relaxor behavior. To investigate this crossover theoretically, we utilize the simple model, based on replica-symmetric solution for disordered Ising model.  We show that our simple model is able to describe qualitatively both the phase diagram and hysteresis loops in TBC.
Based on our previous results in the KTL family of disordered ferroelectrics, we make predictions about dynamical properties of TBC as well as about absence of Curie-Weiss law in paraelectric phase of TBC. Latter fact may be related to the occurrence of Griffiths (para-glass) phase \cite{stef00} in this substance. The schematic phase diagram 
of TBC and similar substances, where a crossover from ferroelectric to glassy behavior occurs, is reported in Fig.1a. 
This phase diagram is more or less standard (see, e.g. Ref. \cite{hoh}) except the region where Griffiths (para - glass) phase is realized. The boundaries of this region may, of course, vary from substance to substance so that para-glass phase may penetrate deeper in paraelectric and/or dipole glass phase.  This is because the different strength of "glassy interaction" between dipoles lead to the formation of glassy clusters (which is a physical mechanism behind Griffiths phase) at higher or lower temperatures. Consequently, these clusters merge into an infinite one (signifying the glassy phase onset) at higher or lower temperatures.

To describe theoretically the ferroelectric - orientational glass crossover behavior in TBC, we should
know what kind of dipoles can order ferroelectrically in this material and how many permissible orientations in the
crystal lattice do they have. For example, it is well-known (see, e.g. \cite{vg90} and references therein) that Li in KTaO$_3$ form impurity dipoles due to off-central position of Li in a host KTaO$_3$ lattice. These dipoles (at $T<T_c(x)$ and at
$x>x_{\rm cr}$, where  $x_{\rm cr}$ is Li critical content) can order ferroelectrically and have six (along [100] kind of directions) permissible orientations in a host lattice. As this information is absent in TBC, we describe the thermodynamic properties of this substance by the simplest possible model of two-orientable dipoles. In such model, to account for interplay between ferroelectric and glassy behavior, we use replica formalism for random Ising model. Namely, we consider the Hamiltonian
\begin{equation}\label{gm1}
{\mathcal H}=-\frac 12\sum_{ij}J_{ij}S_i^zS_j^z-E\sum_iS_i^z,
\end{equation}
where $E\equiv E_z$ is an external electric field (in energy units) and the random interactions $J_{ij}$ between pseudospins $S_i^z$ and $S_j^z$ ($S^z\equiv \pm 1$) are distributed according to Gaussian law
\begin{equation}\label{gm2}
P(J)=\frac{1}{\Delta J\sqrt{2\pi}}\exp\left(-\frac{(J-{\bar J})^2}{2(\Delta J)^2}\right),
\end{equation}
where ${\bar J}$ and $\Delta J$ is, respectively, the mean value (responsible for long-range ferroelectric order formation) and variance (responsible for realization of glassy order parameter and destruction of long-range order) of random interactions. To describe the dependence of TBC thermodynamic characteristics on Pr content $x$, the parameters ${\bar J}$ and $\Delta J$ should be functions of $x$. Both these quantities will be extracted below from the comparison of temperatures $T_c$ (ferroelectric phase transition temperature) and $T_g$ (glassy freezing transition temperature) with their experimental values.

Conventional replica formalism (see, e.g., \cite{koshe,fih}), being applied to Hamiltonian \eqref{gm1} (with respect to Eq. \eqref{gm2}), gives following standard (so-called replica-symmetric \cite{koshe,fih}) equations for long-range $m$ (dimensionless spontaneous polarization) and glassy $q$ order parameters
\begin{widetext}
\begin{eqnarray} \label{mq}
m&=&\frac{1}{\sqrt{2\pi}}\int_{-\infty}^{\infty}e^{-\frac{y^2}{2}}\tanh\left[\frac{E+{\bar J}m+y{\Delta J}\sqrt{q}}{k_BT}\right]dy,\nonumber \\
q&=&\frac{1}{\sqrt{2\pi}}\int_{-\infty}^{\infty}e^{-\frac{y^2}{2}}\tanh^2\left[\frac{E+{\bar J}m+y{\Delta J}\sqrt{q}}{k_BT}\right]dy,
\end{eqnarray}
\end{widetext}
where $k_B$ is Boltzmann constant. These equations (along with corresponding free energy function \cite{fih}) define all equilibrium thermodynamic properties of the system under consideration. We note here the well-known fact that replica-symmetric solution is unstable against replica-symmetry breaking. Under the action of external field $E$ one can draw a borderline, separating the regions of stable and unstable replica-symmetric solutions. In the variables field-temperature, this line is known as de Almeida - Touless (AT) line \cite{at}. Standard (in replica formalism) but lengthy calculation leads to following criterion of replica-symmetric solution stability
\begin{equation}\label{at}
\frac{(\Delta J)^2}{(k_BT)^2}\frac{1}{\sqrt{2\pi}}\int_{-\infty}^{\infty}e^{-\frac{y^2}{2}} \frac{dy}{\cosh^4\left[\frac{E+{\bar J}m+y{\Delta J}\sqrt{q}}{k_BT}\right]}<1.
\end{equation}
To obtain the AT line from Eq. \eqref{at}, we should solve it (the equality in Eq. \eqref{at}) simultaneously with Eqs. \eqref{mq} for order parameters. Our analysis shows that the presence of long-range order parameter $m$ plays a role of additional (to the external one) stabilizing field so that the replica-symmetric solution \eqref{mq} in our case is stable almost everywhere on the $E-T$ plane except very low temperature region. This result is correct for all $x$ except the point $x=0$, where system is ordered. We note that different versions of above replica formalism had been used to describe the disordered dielectrics like proton glasses \cite{bl1} and relaxor ferroelectrics (see Ref. \cite{bl2} and references therein).

First we use the equations \eqref{mq} to calculate the $T-x$ ($T$ is a temperature) phase diagram of the system, i.e. the dependencies $T_c(x)$ and $T_g(x)$, which should be calculated for $E=0$ in Eqs. \eqref{mq}.  Namely, to find $T_c$, we put $m \to 0$ in the first equation \eqref{mq} and expand its integrand at small $m$ up to the first nonvanishing term. This yields
\begin{eqnarray}\label{ryad}
&&\tanh\left[\frac{{\bar J}m+y{\Delta J}\sqrt{q}}{k_BT}\right] \approx \tanh\left[\frac{y{\Delta J}\sqrt{q}}{k_BT}\right] +\nonumber \\
&&+\frac {{\bar J}m}{k_BT}\left(1- \tanh^2\left[\frac{y{\Delta J}\sqrt{q}}{k_BT}\right] \right).
\end{eqnarray}
Substitution of Eq. \eqref{ryad} into the first equation \eqref{mq} yields the vanishing of first term, while the second one gives the equation for $T_c$
\begin{eqnarray}
&&k_BT_c={\bar J}(1-q_0(T_c))\equiv \nonumber \\
&&\equiv \frac{\bar J}{\sqrt{2\pi}}\int_{-\infty}^{\infty}e^{-\frac{y^2}{2}}
\frac{dy}{\cosh^2\left[\frac{y{\Delta J}\sqrt{q}}{k_BT_c}\right] }.\label{tc1}
\end{eqnarray}

Now we pay attention to the fact that since $q\equiv 0$ at $T>T_g$, so that $q(T_c>T_g)=0$, we immediately obtain from Eq. \eqref{tc1} that $k_BT_c={\bar J}$.  Then, putting $q \to 0$ in second Eq. \eqref{mq}, we obtain the expression for $T_g(x)$. Finally, the expressions for both transition temperatures read
\begin{equation}\label{tcg}
k_BT_c={\bar J},\  k_BT_g={\Delta J}.
\end{equation}
To deduce the dependencies of ${\bar J}(x)$ and ${\Delta J}(x)$, we refer to the Fig. 5 from Ref. \cite{mag}, where the experimental dependencies $T_c(x)$ and $T_g(x)$ for TBC are reported. It is seen that there are critical concentrations for glassy $x_g\approx 0.2$ and ferroelectric $x_f\approx 0.8$ orders realization. At all other values of $x$ the dependences $T_c(x)$ and $T_g(x)$ are almost linear. These facts suggest that Pr content dependence of the phase transition temperatures and hence the above parameters of interactions distribution should be chosen in the form of power series in $x$
\begin{widetext}
\begin{equation} \label{par1}
{\bar J}(x)= \left\{
\begin{array}{cc}
  0,                                     &   x\geq x_f \\
 J_0 +a_1x+a_2x^2+...,   &   x<x_f
\end{array}\right. ,\quad
 {\Delta J}(x)=\left\{
\begin{array}{cc}
\Delta J_0+b_1x+b_2x^2+... , &   x>x_g \\
 0,   &   x\leq x_g.
\end{array}\right.
\end{equation}
 \end{widetext}
Here $J_0$ defines the mean interaction between dipoles in TBC at $x=1$, having the physical meaning of a mean electric field (in energy units) acting between the dipoles. In turn, $\Delta J_0$ defines the maximal (at $x=1$) variance of random interactions.  Comparison of Eqs \eqref{tcg} and \eqref{par1} shows the other meaning of parameters $J_0$ and $\Delta J_0$. Namely, $J_0$ defines the mean field (i.e. at $x<x_g$ where the system is almost completely ordered) ferroelectric phase transition temperature, while $\Delta J_0$ measures the maximal (at $x>x_f$) glassy freezing temperature. The situation here is qualitatively similar to that in KTL \cite{hoh}.
Note that dependences \eqref{par1} reflect the physics of the system under consideration: at $x<x_g$ the width of distribution function tends to zero so that Gaussian \eqref{gm2} reduces to
$\delta$ - function. It can be shown \cite{stef97} that this $\delta$ - function generates the well-known mean - field equation $m=\tanh(mJ_0/k_BT)$ for long-range parameter $m$ in ordered Ising model. On the other hand, at $x>x_f$ the mean value ${\bar J}$ equals zero so that the distribution function \eqref{gm2} remains Gaussian but with zero mean, which leads to the situation when the system admits only glassy order parameter, see e.g. Ref. \cite{koshe}.
We note here that while at all $x$ values except critical concentrations, the dependences $T_c(x)$ and $T_g(x)$ are almost linear, near $x_{f,g}$ the whole power series \eqref{par1} come into play. Actually, in Ref.\cite{mag}, the finite values of phase transition temperatures appear abruptly near $x_{f,g}$. Within replica formalism, it is possible to obtain the equations for above critical concentrations $x_{f,g}$. The calculation of $x_{f,g}$, however, involves more subtle analysis, then that based on simple equations \eqref{mq} as below $T_g$ the replica-symmetric solution \eqref{mq} is unstable. The results of such analysis will be published elsewhere. In the Fig.1b, we show the approximation of experimental points (see Fig. 5 of Ref. \cite{mag}) $T_c(x)$ and $T_g(x)$ by linear terms in  \eqref{par1}.
The fitting linear dependencies were obtained in the form:
\begin{eqnarray}\label{apt}
T_c(x)&\approx& 320-29.049x \ (K),  \nonumber \\
T_g(x)&\approx& 157.2+11.57x \ (K).
\end{eqnarray}
From these dependencies we can recover the values $J_0 \approx 320 K$ and $\Delta J_0 \approx 168.8 K$. The comparison of these values shows that $J_0\approx 1.9 \Delta J_0$, which means that mean value of interaction is almost two times larger then its dispersion. Such large ratio $J_0/\Delta J_0$ is actually responsible for the situation when TBC exhibits ferroelectric long-range order for the majority of Pr contents - from $x=0$ to $x=0.8$.
It is seen that the above linear dependencies approximate the experimental points pretty good. This means that our simple replica-symmetric solution captures qualitatively all peculiarities of a crossover between orientational glass and ferroelectricity in TBC.
\begin{figure} [! ht]
\begin{center}
\includegraphics [width=0.49\textwidth]{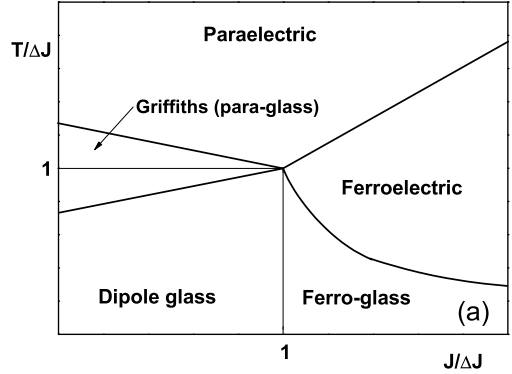}
\includegraphics [width=0.49\textwidth]{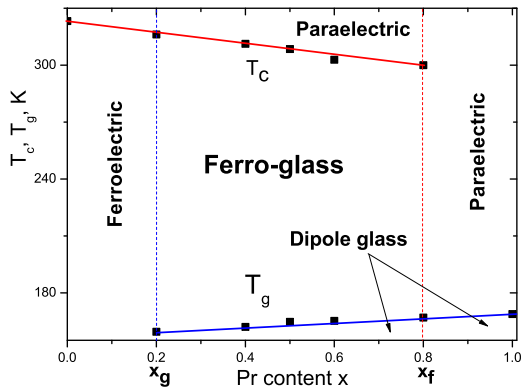}
\end{center}
\caption{The phase diagram of the system under consideration. Panel (a) depicts schematic picture with Griffiths phase shown. Panel (b) reports the 
approximations $T_c(x)=290.95+29.05(1-x)$ K and $T_g(x)\approx 157.2+11.57x$ (full lines) and experimental dependences \cite{mag} $T_c(x)$ and $T_g(x)$ (squares).
Critical $x$ values for glassy $x_g$ and ferroelectric $x_f$ are also shown. }\label{gfi1}
\end{figure}

Next we use the equations \eqref{mq} for the qualitative description of ferroelectric hysteresis $m(E)$ in TBC. Fig.2a reports the qualitative picture of a hysteresis, described by the equations \eqref{mq}. Namely, if $T>T_c(x)$ (paraelectric phase), the curve $m(E)$ is monotonously growing function so that both directions of electric field variation are the same, i.e. we are on the same curve. At $T<T_c(x)$ the $m(E)$ curve acquired "s-shape" with central unstable part with $dm/dE<0$. This unstable part yields the hysteretic behavior, namely jumps at
$E=\pm E_c$, where $E_c$ is a coercive field. These jumps, as usual, generate the hysteresis loop as shown on
Fig.2a. The results of specific calculations for the experimental \cite{mag} temperature $T=280 K$ (corresponding to dimensionless quantity $k_BT/J_0=0.875$) are reported in Fig.2b. It is seen that maximal coercive field is achieved at $x=0$, where the whole system is ferroelectric. As Pr content $x$ grows, the coercive field decays so that at $x$ corresponding to $T_c$ value, the unstable part of the $m(E)$ curve degenerates in a vertical line and $E_c=0$. At $x$ values corresponding to $T>T_c$, we have the paraelectric phase with monotonous $m(E)$ curve similar to that from Fig.2a. We note that saturation polarization diminishes as $x$ grows. This is because at larger $x$ the dilution of the system grows so that there is less (then at $x=0$) ferroelectrically active dipoles, which give smaller saturation polarization.

The comparison of our theoretical hysteresis loops from Fig.2 with experimental ones from Ref. \cite{mag} shows their
different shapes in theory and experiment. While both theory and experiment give the decay of saturation polarization as $x$ grows, the coercive field in the experiment and theory have different behaviors. Namely, the experimental coercive field depends on $x$ only weakly, while the theoretical one strongly. Latter fact gives the different shapes of theoretical and experimental hysteresis loops. To calculate the above shape more precisely, the consideration of repolarization of domain structure of TBC in the external electric field is necessary. This problem requires experimental investigations of character of ferroelectric domain structure in TBC. The theoretical approach dealing with domain structure characteristics in disordered ferroelectrics, have been put forward earlier \cite{mu08}. Other way of (less physical) calculation of hysteresis loops in TBC is to use Preisach model \cite{preisach} - like approach (see e.g. \cite{delt} and references therein), where the loops $m(E)$ from Fig. 2b (we recollect that these are actually the solutions of Eqs. \eqref{mq} at $T<T_c(x)$) can be considered as elementary hysterons. As those hysterons are functions of $T$ and $x$, the resulting loop would also depend on these parameters.

\begin{figure} [! ht]
\begin{center}
\includegraphics [width=0.49\textwidth]{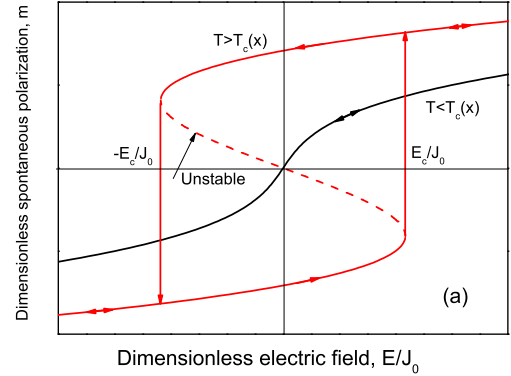}
\includegraphics [width=0.49\textwidth]{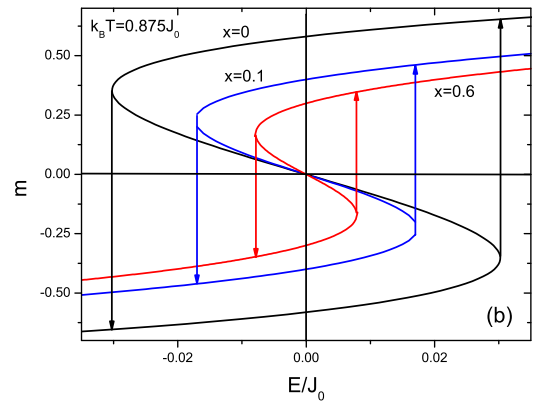}
\end{center}
\caption{The theoretical hysteresis loops $m(E/J_0)$, obtained from solution of equations of Eqs. \eqref{mq}. Panel (a) shows qualitative situation. Dashed line corresponds to the unstable part of the $m(E/J_0)$ curve at $T<T_c(x)$. Arrows show the possible directions of sweep around hysteresis loop. The coercive field $E_c$ is also shown. Panel (b) reports quantitative solution for $k_BT=0.875 J_0$ and different $x$. }\label{gfi2}
\end{figure}

To conclude, in the present paper we have shown that tungsten bronze compound (TBC) Ba$_2$Pr$_x$Nd$_{1-x}$FeNb$_4$O$_{15}$ undergoes a crossover between ferroelectric and orientational (dipole) glass phases rather then exhibits "classical" (i.e. that adopted in the literature) relaxor behaviour. To demonstrate that, we use the replica-symmetric solution of disordered Ising model which permits to derive the dependences of phase transition temperatures $T_c$ and $T_g$ on Pr content $x$. We note here that the above replica formalism permits to obtain the equations for critical concentrations for ferroelectric $x_f$ and glassy order $x_g$ appearance.  This issue can be elaborated considering the nonequilibrium (like frequency dependent dielectric susceptibility) properties of the system under consideration; it is out of frames of present paper.

The same replica-symmetric equations \eqref{mq} have been used to calculate the dependence $m(E)$, determining qualitatively the ferroelectric hysteresis in the system. We have shown that hysteresis behaves according to the system phase diagram. Namely, as (at given $x$) $T$ approaches $T_c$ or $x \to x_f$, the loops monotonously shrink, giving smaller coercive field values. At $T=T_c$ or $x = x_f$ the loop has zero width and coercive field $E_c=0$.
Although our theory describes the decay of saturation polarization at Pr content growth, we were not able to explain the enhancement of $E_c$ (the "width" of hysteresis loop) in TBC at $x=0.6$, see Fig. 6 of Ref. \cite{mag}. The explanation of this puzzling behavior requires (even on the level of replica-symmetric solution \eqref{mq}) more exquisite approaches (then above simple model) like supposition that there are additional defects (interacting with ferroelectrically active dipoles) in the system such that at certain $x$ (e.g. at $x=0.6$ in Ref. \cite{mag}) they enhance the spontaneous polarization and consequently coercive field. The experimental fact \cite{mag} that paraelectric phase of TBC does not follow the Curie-Weiss behavior, can be explained in terms of Griffiths (paraglass) phase realization in the substance \cite{stef00}, see Fig.1a. Although the dynamics of order parameters cannot be calculated with the help of above replica-symmetric solution, its qualitative description can be made similar to that in ferroelectric polymers \cite{boz03}. Such calculations are out of frames of a present work and will be published elsewhere. We emphasize once more that to elucidate the interplay between ferroelectric and orientational glass behavior in nonperovskite disordered ferroelectrics, additional experimental investigations are highly desirable.

This work was supported in part by Opole University Intramural Grant (Badania Statutowe).

\end{document}